\documentclass[a4paper]{article}
\usepackage{graphics}
\usepackage[dvips]{graphicx}
\usepackage{epsfig}
\usepackage{amsmath}
\usepackage{amsfonts}
\usepackage{amssymb}
\usepackage{wasysym}
\usepackage{color}
\usepackage{ulem}    

\def\be{\begin{equation}}
\def\ee{\end{equation}}
\def\bfi{\begin{figure}}
\def\efi{\end{figure}}
\def\bea{\begin{eqnarray}}
\def\eea{\end{eqnarray}}

\begin{document}

\title{Activity-dependent neuronal model on complex networks}

\author{Lucilla de Arcangelis$^{1}$ and Hans J. Herrmann$^{2,3}$}

\maketitle

\centerline{\it
$^{1}$ Department of Information Engineering, Second University of Naples, 
Aversa (CE), Italy}

\centerline{\it
$^{2}$  Institute Computational Physics for Engineering Materials,
ETH, Z\"urich, CH}

\centerline{\it
$^{3}$ Departamento de F\'{\i}sica, Universidade Federal do Cear\'a,
60451-970 Fortaleza, Cear\'a, Brazil
}


\begin{abstract}
Neuronal avalanches are a novel mode of activity in neuronal networks,
experimentally found in vitro and in vivo, and exhibit a robust critical
behaviour: These avalanches are characterized by a power law distribution for
the size and duration, features found in other problems in the context of the
physics of complex systems. 
We present a recent model inspired in self-organized criticality, which
consists of an electrical network with threshold firing, refractory period and
activity-dependent synaptic plasticity. The model reproduces the
critical behaviour of the
distribution of avalanche sizes and durations measured experimentally.
Moreover, the power spectra of the electrical signal reproduce very robustly
the power law behaviour found in human electroencephalogram (EEG) spectra.
We implement this model on a variety of complex networks, i.e. regular,
small-world and scale-free and verify the robustness of the critical behaviour.

\end{abstract}
\section{Introduction}

The activity in neuronal networks consists in one or more action potentials in
a single neuron or an ensemble of neurons. The first case is typical for  small
networks, as some experimental systems {\it in vitro}, where isolated spikes
can
be observed. The presence of a number of action potentials in an ensemble of
neurons not always is a consequence of an external stimulus. Neuronal systems
exhibit an intense spontaneous activity, known since a long time, whose
relation
with the response to stimulation is not fully understood yet.
It is however well established that spontaneous activity cannot be simply
reduced to a background noise uncorrelated to the system response. Indeed,
experimental results for the cat visual cortex (\cite{Arie}) have shown that the
intensity of the response to an external stimulus is roughly proportional to
the
intensity of the spontaneous activity state of the system when the stimulus is
applied. The variability in the response provided to the repeated application
of the same stimulus is therefore caused by the different levels of ongoing
activity.  A similar analysis has been performed at the intracellular level
on the same system, confirming that the spatio-temporal structure of the
spontaneous activity influences the response signal (\cite{Azou}).

The typical form of spontaneous activity consists in the almost synchronous
emission of action potentials in a large number of neurons, followed by
periods of substantial inactivity. These high activity events, named bursts,
are observed both during development and in mature systems and can last from a
few to several hundreds milliseconds. Conversely, the quiet periods can last
seconds and have been attributed to a variety of mechanisms:
The decrease in the available neurotransmitter (\cite{ste,sta}); the presence of
an inhibitory factor leading to a disabilitation of the neurotransmitter
release (\cite{ste,sta}); the inactivation, or remodulation of the response, of
the glutamate receptors  (\cite{mae}).  An alternative form of
temporal organization is slow oscillations  between high activity and low
activity states with a typical frequency of 0.3 - 1 Hz.
The temporal organization of this spontaneous activity has been characterized
by the distribution of inter-times, i.e. the temporal intervals between
successive bursts or successive spikes (\cite{seg}).

In 2003 Beggs and Plenz have identified a novel form of spontaneous activity,
neuronal avalanches (\cite{bp03,bp04}). Coronal slices of rat
somatosensory cortex were placed onto a $8\times8$ multielectrode array (MEA) and
spontaneous activity was induced by bath perfusion with the glutamate receptor
agonist NMDA in combination with a dopamine receptor agonist.
The intrinsic activity of the system was monitored by measuring the potential
at
each electrode. This local field potential (LFP) integrates the electrical
activity
of neurons placed in the region surrounding the electrode: negative peaks in
the LFP
measure the influx of positive ions and therefore the cumulative membrane
potential
variation of the neurons in the region. Experimental data show that before 6
days in vitro activity is mainly composed of sparse activations but during the
second week
simultaneous activations occur in several electrodes. The novel idea was to
examine
this electrophysiological signal on a finer temporal scale, which was able to
evidence a complex spatio-temporal structure. Indeed, activity starting at one
electrode may involve more, non necessarily neighbouring, electrodes. Binning
time in cells of duration $\delta t$, allows to create a
spatio-temporal grid reporting the  active electrodes in each temporal cell.
A neuronal avalanche is therefore defined as a sequence of successively active
electrodes between two temporal bins with no activity. The total number of
active electrodes, or alternatively the sum of all LFPs, is defined as the size
$s$ of an avalanche and the time interval with ongoing activity as its duration
$T$.

The striking result is that both size and duration have no characteristic
value,
i.e. their distributions exhibit a power law behaviour. The analysis at a finer
temporal scale is then able to enlighten the non synchronous character of the
bursts. The exponents of these power law distributions depend on the choice for
the temporal bin  $\delta t$. Indeed larger  bins make active electrodes
belonging to different avalanches to merge into the same larger event, leading
to a smaller exponent. In order to identify the appropriate value of  $\delta
t$,
Beggs and Plenz verified that if  $\delta t$ is equal to the average value of the
time delay between two successive LFPs in the culture, the
exponent does not depend any longer on the specific culture. They were then
able to identify the universal scaling behaviour

\begin{eqnarray}
P(s)\propto s^{-\alpha} \quad {\rm with} \quad
\alpha=1.5\pm 0.1 \nonumber   \\
P(T)\propto T^{-\beta} \quad {\rm with} \quad \beta=2.0\pm 0.1.
\label{ple}
\end{eqnarray}

The power law behaviour for the size distribution is followed by an exponential
cutoff due to the finite size of the system, whereas for the duration
distribution
it extends over about one decade and the exponential cutoff sets in at about
10ms.

The results {\it in vitro} have been confirmed by extended studies
{\it in vivo} on anaesthetized rats during development (\cite{gp08}) and awake
reshus monkeys (\cite{pp09}). Spontaneous neuronal activity recorded by MEA
placed in the rat cortical layer 2/3 at the beginning and the end of the second
week postnatal, shows higher frequency (up to 100Hz) oscillations nested into
lower frequency  (4-15Hz) oscillations. At the end of the first week postnatal,
bursts start to organize into high frequency oscillations and become more
synchronized during the second week.
Synchronous activity in the bursts exhibits the
same scaling behaviour found for neuronal avalanches {\it in vitro}
(Eq.s\ref{ple}). This similarity between {\it in vitro} and {\it in vivo}
experiments support the idea that the emergence of nested oscillations reflects
the development of layer 2/3 in the cortex.
Ongoing activity measured in the primary motor and premotor
areas of two awake monkeys, sitting with no behavioural task, nor under
particular stimulus, exhibits also neuronal avalanches. Their
organization is independent of the detection threshold and exhibits
scale invariance. Power laws for the size and duration distributions confirm
the scaling behaviour in Eq.\ref{ple} and suggest that in large neuronal
networks a wide variety of avalanche sizes is possible, including clusters
percolating throughout the system. This indicates that the largest cluster
is solely controlled by the system size and not by the dynamics.
This result also generalizes avalanche dynamics across
species and different cortical areas. Criticality can be therefore considered
as a generic property of spontaneous cortical activity, which may indicate that
networks with a larger  response repertoire were selected over
others throughout evolution. A flexible spontaneous activity could then
underlie and optimize important cortical functions as learning and memory.

The investigation on the spontaneous activity has been performed also for
dissociated neurons from different networks as rat hippocampal neurons
(\cite{maz}), rat embryos (\cite{pas}) or leech ganglia (\cite{maz}).
Neurons  are mechanically dissociated by trituration through fine-tipped
pipettes and placed onto  a MEA, pre-coated with adhesion promoting molecules,
in a nutrient medium. Under fixed conditions of humidity and temperature,
neurons start to develop a network of synaptic connections and, after a
variable period {\it in vitro}, exhibit spontaneous electrical activity.
The electrodes of the MEA in these experiments record the spikes, rather than
the LFPs, due to individual neurons attached to them. As a consequence, the
temporal scale for the data analysis has to take into account this difference
in order to properly characterize the neuronal response. Choosing
the average inter-spike time at a single electrode as the temporal scale
for data binning, the spontaneous activity is monitored during the development
and in mature cultures. Different behaviours are observed. Only those systems
exhibiting a medium level of synchronization between random spikes and
synchronized bursts exhibit critical behaviour. For those cultures
the scaling behaviour is very robust and in agreement with Eq.\ref{ple}.
In particular, the emergence of a critical state has been found
to be strongly related to the aging of the system, namely after the first few
weeks
{\it in vitro}, where the behaviour of the system is subcritical, some
cultures may
self-organize and reach the critical state as they mature (\cite{pas}).

In real brain neurons are known to be able to develop an extremely high
number of
connections with other neurons, that is a single cell body may receive inputs
from even a hundred thousand presynaptic neurons. One of the most
fascinating questions is how an
ensemble of living neurons self-organizes, developing connections
to give origin to a highly complex system. The dynamics underlying this
process might be driven both by the aim of realizing a well connected network
leading to efficient information transmission, and the energetic cost of
establishing very long connections. The morphological characterization of a
neuronal network grown {\it in vitro} has been studied
(\cite{shefi}) by
monitoring the development of neurites in an ensemble of few hundred neurons
from the frontal ganglion of adult locusts. After few days the cultured
neurons have developed an
elaborated network with hundreds of connections, whose
morphology and topology
has been analyzed by mapping it onto a connected graph. The short path length
and the high clustering coefficient measured indicate that the network
belongs
to the category of small-world  networks (\cite{watts}), interpolating between
regular and random networks. In classical small-world networks the majority of sites have
a number of connections close to the average value in the network. Real
neuronal networks behave quite differently, since neurons with quite diverse
number of connections are observed. Indeed, the properties of the 
functionality network have been measured experimentally  in human 
adults (\cite{chia2}). Functional magnetic resonance imaging has 
shown that this network has universal scale free properties, namely it
exhibits a distribution of out-going connection number, $k_{out}$,
which follows a power law, i.e. $n(k_{out})\propto k_{out}^{-2}$,
independent of the different  tasks performed by the patients. This behaviour
suggests that in the network few neurons are highly connected and act as hubs
with respect to information transmission.
Small world features have been also measured for functionality networks in
healthy
humans, whereas they are not present in patients affected by neurological
diseases: Alzheimer
patients have longer path lengths (as in regular networks) (\cite{stam}) 
whereas schizophrenic patients
show a more random architecture of the underlying network (\cite{rubi}). 
Epileptic patients exhibit a
more ordered neuronal network during seizures (\cite{pont}), whereas brain tumour
patients a more random one (\cite{bart}).

\section{The model}

\subsection{Connectivity networks}

The first step to develop a model simulating neuronal dynamics is the choice
of the specific network of connections.
The simplest choice is  a regular lattice, i.e. a square lattice for a
two-dimensional system. However, following recent experimental results, 
we allow neurons to develop long range  connections: Starting from a regular
lattice, a  small fraction of bonds, from 0 to $10 \%$, is rewired, namely
one of the two connected neurons is chosen at random in the system. This
procedure originates long range connections and gives rise
to a small world network (\cite{watts,shefi}), 
which more realistically reproduces the connections in the real brain.

In a small world networks the number of connections for different neurons
is close to an average number. In order to reproduce the experimental data
on the connectivity distribution in functionality networks, we implement also
scale free networks. 
More precisely, we set $N$ neurons at random positions in two-dimensional space
and to each neuron we assign an out-going connectivity degree,
$k_{out}$, according to the distribution measured
by fMRI measurements of ongoing activity in
humans (\cite{chia2}). 
Each neuron has a degree equal to a random number between
$k^{min}_{out}=2$ and  $k^{max}_{out}=100$ according
to the probability distribution $n(k_{out})\propto k_{out}^{-2}$.
The two neurons are chosen according to a distance dependent probability,
$p(r)\propto e^{-r/5<r>}$, where $r$ is their spatial distance (\cite{roe}).

In order to consider a network with both features, small-world and
scale-free, we also implement
the Apollonian network. This has been recently introduced (\cite{apo}) in a 
simple deterministic
version starting from the problem of space-filling packing of spheres according
to the ancient Greek mathematician Apollonius of Perga. In its classical
version
the network associated to the packing gives a triangulation that physically
corresponds to the force network of the sphere packing.
One starts with the zero-th order triangle of corners
$P_1, P_2, P_3$, places a fourth site $P_4$ in the centre of the triangle
and connects it to the three corners ($n=0$).
This operation will divide the original triangle in three smaller
ones, having in common the central site. The iteration $n=1$ proceeds
placing one more site in the centre of each small triangle and connecting
it to the corners (Fig.1). At each iteration $n$, going from 0 to $N$,
the number of sites increases by a factor 3 and the coordination
of each already existing site by a factor 2.
More precisely, at generation $N$ there are
$$m(k,N)=3^N, 3^{N-1},3^{N-2},\ldots,3^2,3,1,3$$
vertices, with connectivity degree
$$k(N)=3,3\times 2,3\times 2^2,\ldots,3\times 2^{N-1},3\times 2^N,2^{N+1}+1$$
respectively, where the two last values correspond to the site $P_4$ and the
three corners  $P_1, P_2, P_3$.
The maximum connectivity value then is the one of the
very central site $P_4$, $k_{max}=3\times2^N$, whereas the sites inserted at
the $N$-th iteration will have the lowest connectivity 3.

\begin{figure}
\centering
\includegraphics[width=6cm]{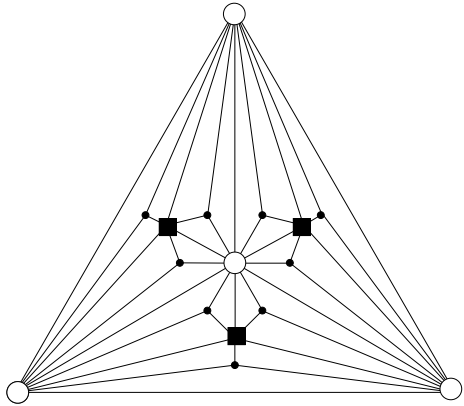}
\caption{
Apollonian network for $N=2$: iterations $n=0,1,2$
are symbols $ \bigcirc , \blacksquare , \bullet$, respectively.
} \label{Fig1}
\end{figure}

The important property of the Apollonian network is that it is scale-free.
In fact, it has been shown (\cite{apo}) that the discrete cumulative 
distribution of connectivity degrees
$P(k)=\sum_{k'\ge k} m(k,N)/N_N$, where $N_N=3+(3^{(N+1)}-1)/2$
is the total number
of sites at generation $N$, can be fitted by a power law. More precisely,
$P(k)\propto k^{1-\gamma}$, with $\gamma =\ln 3 /\ln 2 \sim 1.585$. Moreover
the network
has small-world features. This implies (\cite{watts}) that the average length of
the shortest
path $l$ behaves as in random networks and grows slower than any positive
power of $N$,
i.e. $l\propto (\ln N)^{3/4}$. Furthermore the clustering coefficient $C$ is
very high
as in regular networks ($C=1$) and contrary to random networks.
For the Apollonian network $C$ has been found to be equal to $0.828$ in the
limit of large $N$.
On this basis, the Apollonian network appears to have all the new features
that we would like to investigate:
small-world property found experimentally (\cite{shefi}) and possibility
of a very high connectivity degree (scale-free). Moreover it also presents
sites
connecting bonds of all lengths. Also this last feature can be found
in real neuronal networks, where the length of an axon connecting the
pre-synaptic with the post-synaptic
neuron can vary over several orders of magnitude, from $\mu m$s to $cm$s.
Finally, most studies in the literature consider the case of a fully connected
network, where each neuron is connected to every other neuron. Even if not 
completely realistic, we consider also this last case.

\subsection{Neuronal dynamics}
We here discuss a neuronal network model
based on self-organized criticality ideas (\cite{bak}). 
The model implements several physiological
properties of real neurons: a continuous membrane potential, firing at
threshold, synaptic plasticity and pruning.
In order to define the model we need to specify the 
behaviour of the single neuron
under different conditions, the dynamics then determines the system 
behaviour (\cite{dea1,dea2,dea3}).
We consider $N$ neurons at the nodes of the chosen network, 
characterized by their potential $v_i$. The neuron positions will then be
ordered in space for regular lattices and small world networks, 
organized in a hierarchical manner for the Apollonian network and randomly
chosen in two dimensions for the scale-free and fully connected networks.
Once the network of output connections is established, we identify the resulting
degree of in-connections, $k_{in_j}$, for each neuron $j$.
To each synaptic connection we assign an
initial random strength $g_{ij}$, where $g_{ij}\neq g_{ji}$, and to each neuron
randomly either
an excitatory or an inhibitory character, with a fraction $p_{in}$ of inhibitory
synapses. Whenever at time $t$ the
value of the potential at a site $i$ is above a certain threshold
$v_i \geq v_{\rm max}$, the neuron sends action potentials
which arrive to each of the $k_{out_i}$ pre-synaptic buttons and
lead to a total production
of neurotransmitter proportional to $v_i$. As a consequence, the
total charge that could enter into connected neurons is proportional to 
$v_i k_{out_i}$.
Each of them receives charge in proportion to the strength of the synapses
$g_{ij}$
\be
v_j(t+1)=v_j(t)\pm \frac{v_i(t) k_{out_i}}{k_{in_j}}\frac{g_{ij}(t)}{\sum_k g_{ik}(t)}
\label{prop}
\ee
where the sum is extended to all out-going
connections of $i$. In Eq.(2) the membrane potential variation is obtained
by dividing the received charge by
the surface of the soma of the post-synaptic neuron, proportional
to the number of  in-going terminals $k_{in_j}$. The plus or minus sign in
Eq.(2)  is for excitatory or inhibitory synapses, respectively.
In regular networks neurons have the same number of ingoing and
outgoing connections, therefore Eq.(1) reduces to the simpler expression
$v_j(t+1)=v_j(t)\pm v_i(t) \frac{g_{ij}(t)}{\sum_k g_{ik}(t)}$.
The same consideration holds for small world networks.

The firing rate of real
neurons is limited by the refractory period, i.e. the brief period after the
generation of an action potential during which a second action potential is
difficult or impossible to elicit. The practical implication of refractory
periods is that the action potential does not propagate back toward the
initiation point and therefore is not allowed to reverberate between the cell
body and the synapse. In our model, once a neuron fires, it remains quiescent
for one time step and it is therefore unable to accept charge from firing
neighbours. This ingredient indeed turns out to be crucial for a controlled
functioning of our numerical model. In this way an avalanche of charges can
propagate far from the input through the system.
The initial values of the neuron potentials are uniformly distributed 
random numbers and the value of $v_{max}$ is fixed equal to 6 in all
simulations.
Moreover, a small fraction ($10\%$) of neurons is chosen to be output sites,
i.e. an open boundary, with a
zero fixed potential, playing the role of sinks for the charge. They model
neurons
connected to neurons not belonging to the slice and avoid that an excess
to charge influx would lead to supercritical behaviour.
Each time neuronal activity stops in the network, an external stimulus is
necessary to trigger further activity, which therefore mimics the
nutrients from the bath needed to keep a real neuronal network alive.
This stimulus consists in increasing the potential of a random neuron by a
random quantity uniformly distributed between 0 and $v_{max}$.

During the propagation of an avalanche according to Eq. (2), we identify the
bonds connecting two successively active neurons, namely neurons whose activity
is correlated. The strength of their connections is increased proportionally to
the
activity of the synapse, namely the
membrane potential variation of the post-synaptic neuron induced by the
presynaptic neuron
\be
g_{ij}(t+1) =g_{ij}(t) +\alpha i_{ij}(t)
\ee
where $i_{ij}(t)$ is the current through that synaptic connection and
$\alpha$ a dimensionless parameter.
Once an avalanche of firings comes to an end, the strength of all inactive synapses
is
reduced by the average strength increase per bond
\be
\Delta g = \sum_{ij, t} \delta g_{ij} (t)/ N_a
\ee
where $N_a$ is the number of bonds active in the previous avalanche. Here $\alpha$ is the only
parameter controlling both the strengthening and the weakening rule in the
Hebbian plasticity and represents the ensemble of
all possible physiological factors influencing synaptic plasticity.
By implementing these rules, our neuronal network "memorizes" the  most used
paths
of discharge by increasing their strength, whereas the less solicited synapses
slowly
atrophy. Indeed, once the strength of a bond is below an assigned small value
$g_t=10^{-4}$, we remove it, i.e. set its strength equal to zero, 
which corresponds to the so-called pruning.

\begin{figure}
\vskip+1cm
\centering
\includegraphics[width=7cm]{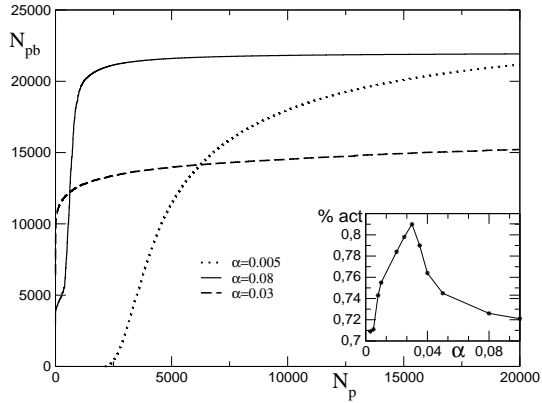}
\caption{ Average number of pruned bonds $N_{pb}$ as function of the number of
external stimuli $N_p$
for a square lattice of linear size $L = 100$,
 equal initial conductances
and different values of $\alpha$.
In the inset we show the asymptotic
value of the percentage of surviving bonds as function of $\alpha$.
}
 \label{Fig2}
 \end{figure}

We implement synaptic plasticity rules during a series of $N_p$ stimuli in
order
to modify the synaptic strengths, initially set at random. In this way we do
not impose a strength configuration but let the system activity tune their
values.
Once a percentage of bonds is pruned, we stop plastic
adaptation and we perform our measurements, by applying a new sequence of 
stimuli without modifying the synaptic strengths. The extension of the plastic
adaptation procedure then represents the level of experience, or {\it age}, of the
system, whose response we monitor over a time-scale much shorter than the one
needed for structural adaptation.
All data presented in this manuscript are averaged over long temporal sequences
in several initial network configurations. More precisely, for regular and
small-world networks we average data on 10 different initial configurations
with a sequence of 10000 avalanches per configuration. On the Apollonian network we
average over 100 different initial configurations and a sequence of 30000
avalanches per configuration. For scale free and fully connected networks we
average over 60 different initial configurations and a sequence of 50000
stimulations per configuration. 

\section{Pruning}
The total number of pruned bonds at the end of each avalanche,
$N_{pb}$,  in general  depends on the initial
conductance $g_0$, therefore it is interesting to investigate the two cases of
either all initial conductances equal to 0.25, or being uniformly 
distributed between 0 and 1.
First the case of equal initial conductances is analysed. 
The strength of the
parameter $\alpha$, controlling both the increase and decrease of synaptic
strength, determines the plasticity  dynamics in the network.
This homeostatic mechanism implies that
the more the system learns strengthening the used synapses,
the more the unused connections will weaken.
For large values of $\alpha$ the system strengthens more
intensively the synapses  carrying current but also very
rapidly prunes the
less used connections, reaching after a short transient a plateau where it
prunes very few bonds. On the contrary, for small values of $\alpha$
the system takes more time to initiate the pruning  process and
slowly reaches a plateau (Fig. 2).
The inset of the figure shows the asymptotic value of
the fraction of surviving bonds, calculated as the total number of bonds in the
unpruned network  minus the asymptotic number of pruned bonds,
as function of $\alpha$.
The number of unpruned bonds asymptotically reaches its largest
value at the value $\alpha\simeq 0.03$ for different networks.
This could be interpreted as an optimal value for the
system with respect to plastic adaptation.


For the Apollonian network
it is interesting to investigate if pruning acts in the same way 
on bonds created at different iterations $n, n=0,\ldots,N$,
or rather tends to eliminate bonds of some particular iteration.
The probability
to prune bonds of different $n$ is evaluated, that is the number of
pruned bonds over the
total number of bonds for each iteration stage, as function of the number
of applied stimuli. Fig.3 shows that the
plateau is reached at about the same value of $N_p$ and the shape of the
curve is similar for each $n$. However the probability to prune bonds
with large $n$ is higher: These are the bonds created in the last iterations
and therefore embedded in the interior of the network. This suggests that the
most active bonds are the long range ones (small $n$), which therefore support
most of the information transport through the network. It is also interesting
to notice that, since the total number of bonds depends exponentially on $n$,
the gaps between the asymptotic values of the probability for
successive generations depend exponentially on $n$. 
In the inset of Fig.3 we show
the asymptotic number of pruned bonds per generation on a semi-log scale, this
quantity is well fitted by the exponential behaviour
$N_{pb}\simeq \exp {n}$.

\begin{figure}
\centering
\includegraphics[width=7cm,angle=-90]{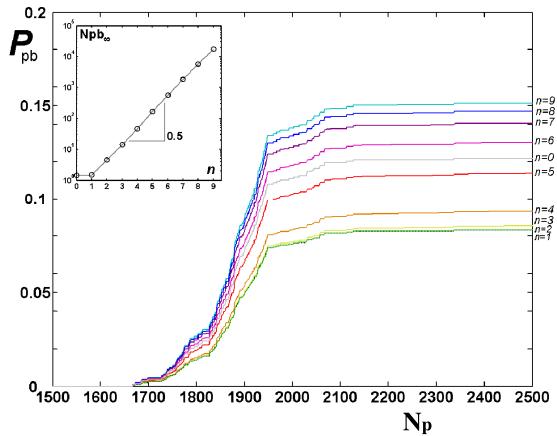}
\caption{Probability of pruning for bonds of different
iterations $n$ of Apollonian networks, from bottom $n=0$ to top $n=9$,
as function of the number of external stimuli $N_p$
for equal initial synaptic strengths. In the
inset, the asymptotic $N_{pb}$ (after 5000 stimuli) is shown as 
function of $n$ with
the exponential fit $N_{pb}\simeq \exp {0.5 n}$.
} \label{Fig3}
\end{figure}

The same analysis has been performed for random initial conductances between
0 and 1. The results are similar to the previous case. It can be noticed that
pruning starts already
at $N_p=1$, since conductances close to zero are present, and the plateau is
reached after about 1000 stimuli. The value of $\alpha$
which now optimizes the number of active bonds is about 0.030 also for the
Apollonian network.
In this case,  the pruning behaviour for different iterations is similar to the
previous case, with the pruning probability exponentially increasing with $n$,
as $N_{pb}\simeq \exp {n}$.

The effect of pruning on the connectivity degree of the network is an
interesting quantity to monitor on scale free networks.
On Apollonian networks we evaluate the number of sites with a number of
outgoing connections
$k_{out}$ as function of $k_{out}$ in the initial network and after application
of a given number of external stimuli (Fig.4). 
After the application of
few external stimuli, i.e. for a short plastic adaptation, the distribution
$n(k_{out})$ shows the same scaling behaviour of the original Apollonian network.
As the pruning process goes on, sites exhibit varying connectivity degree and new
values of $k_{out}$
appear. The result is that  the scaling behaviour is progressively lost, as
well as the scale-free character
of the network, since there is a generalized decrease of the connectivity
in the network.

\begin{figure}
\vskip+1cm
\includegraphics[width=8cm]{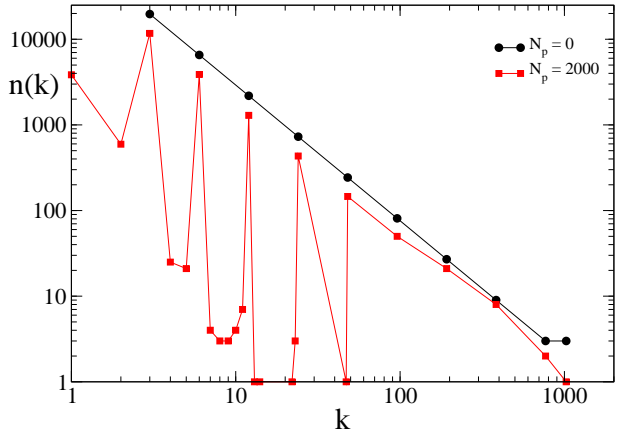}
\caption{Connectivity degree distribution $n(k_{out})$ at
different pruning stages $N_p$ for Apollonian networks with
equal initial synaptic strengths and
$\alpha=0.020$. As soon as pruning starts to eliminate bonds,
new connectivity degrees appear, not present in the original network.
Conversely, two out of the three corner sites, which for the generation
$N=9$ have initially a
connectivity degree 1025, may loose bonds because of pruning and,
as a result, $n(1025)=1$. 
}
\label{Fig.4}
\end{figure}

\section{Avalanche statistics}

After "aging" the system applying plasticity rules during $N_p$ external
stimuli, we submit the
system to a new sequence of stimuli with no modification of synaptic
strengths. The response of the system to this second sequence models the
spontaneous activity of a trained
neuronal network with a given level of experience. We analyse this activity
by measuring the avalanche size distribution $n(s)$ and
the time duration distribution $n(T)$.

\begin{figure}
\vskip+1cm
\includegraphics[width=8cm]{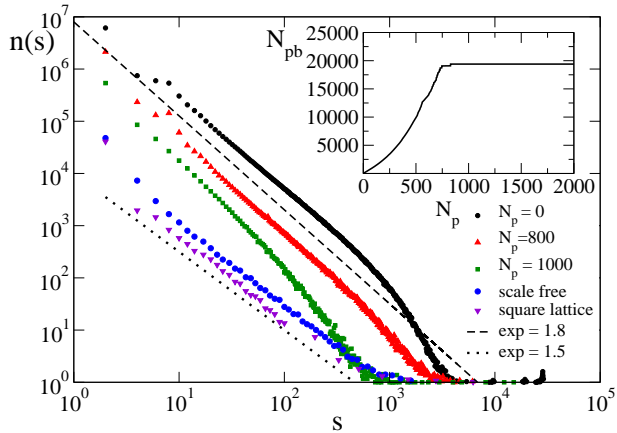}
\caption{Avalanche size distribution for different networks with  $p_{in}=0.05$:
the square lattice ($N = 10^6$, $\alpha=0.03$); the scale free network
($N=4000$) and the Apollonian network for different values
of $N_p$ (9th generation with $\alpha =0.030$). Initial synaptic strengths are
randomly distributed. Data are logarithmically binned.
In the inset, the corresponding behaviour of the number of pruned bonds for
the Apollonian network is shown.
}
\label{Fig.5}
\end{figure}

The avalanche size distribution $n(s)$ consistently exhibits power law
behaviour for different values of model parameters.
Fig.5 shows the avalanche size distribution for different networks and 
values of $N_p$,
including also the case $N_p=0$ (no plasticity adaptation), for random initial
conductances and the optimal value of $\alpha$.
The value of the exponent is obtained by regression of the
log-binned data and found to be $\sigma=1.5\pm 0.1$ for all networks, 
except the Apollonian network where $\sigma=1.8\pm 0.2$.
The exponent is  stable with
respect to variations of the parameters  for both equal and random
initial conductances. More accurate methods, as maximum likelihood
fitting, should verify the stability of these values (\cite{clau}).

It is interesting to stress the importance of noise:
Indeed, by applying the external stimulation not at random but at a fixed 
neuron,  the scaling exponent becomes $\sigma=1.2\pm 0.1$ (\cite{dea1}).
We notice that, for fixed size $s$, increasing $N_p$ decreases the number
of avalanches of that size, suggesting that strong plasticity remodelling
decreases activity. The exponent appears to be independent of  $N_p$, as long
as the number of pruned bonds,
$N_{pb}$, is far from the plateau (see inset in Fig.5). Similar results are
found for equal initial conductances. 
The dependence of the critical behaviour on synaptic strengths has been
recently investigated in networks of integrate-and-fire neurons
(\cite{lev}).
The value of the exponent is  compatible within error bars with the value found in the
experiments of Beggs and Plenz (\cite{bp03}), $1.5\pm 0.4$.
However, one has to notice that experimental results for neuronal avalanches
were obtained for local field potentials, i.e. the underlying events
correspond to local population spikes, whereas the numerical events are
single neuronal spikes.
The slightly larger value of the exponent, found on the Apollonian network,
suggests that the peculiar hierachical structure of the network may reduce the
probability of very large avalanches but does not change substantially the
electrical activity.
For larger $N_p$, the distribution exhibits an increase in
the scaling exponent and finally looses the scaling behaviour for very large
$N_p$ values, in the plateau regime for the number of pruned bonds.

\begin{figure}
\vskip+1cm
\includegraphics[width=8cm]{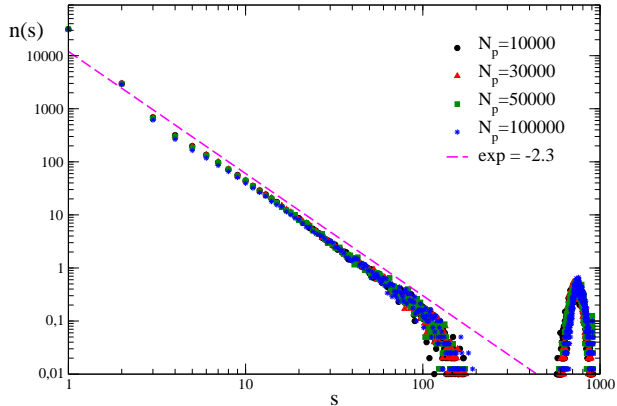}
\caption{Avalanche size distribution for 100 configurations of fully connected
networks with N=1000 neurons with $p_{in}=0.05$.
The different curves correspond to different durations of the plastic
adaptation period $N_p$.}
\label{Fig6}       
\end{figure}

In order to investigate  the role of plastic modifications
on the production of very large avalanches, simulations are performed for
fully
connected networks which undergo plastic adaptation routines of different length.
All networks  exhibit supercritical behaviour, namely an excess of very large
avalanches, due to
the high level of connectivity in the system (Fig.6). 
Very large avalanches
involve almost all neurons and their large number hinders pruning,
namely
there are only very few synapses in the system repeatedly inactive which progressively
weaken and atrophy. This behaviour is independent of the extension of the
plastic adaptation. No pruning is observed even following the application of
hundred thousand stimuli. Very large avalanches therefore seem to be
sustained by the high connectivity in the system and apparently do not depend
on the synaptic strengths.
The analysis of the effect of pruning on very large avalanches confirms this
observation. Plastic adaptation of different duration is now applied to scale
free networks, leading to pruning of synapses. Supercritical behaviour, that
appears
in the unpruned networks, survives when only few percentage of
the synapses is removed. Conversely, a more extended pruning strongly affects
 connectivity and hampers the development of very large avalanches.

At time $t=0$ a neuron is activated by an external stimulus initiating the
avalanche. This will continue until no neuron is at or above threshold. The
number of avalanches lasting a time $T$, $n(T)$, as function of $T$ also 
exhibits
power law behaviour (Fig.7) followed by an exponential cut-off.
The scaling exponent is found to be $\tau=2.1\pm 0.2$  for all networks and equal
and random initial conductances. Only for the fully connected networks
the distribution exhibits a bump at long durations, due to the excess of large
avalanches which all contribute to the tail of the distribution.
The value  of the exponent is found to be stable with respect to different parameters,
provided that the number of pruned bonds $N_{pb}$ is lower than
the plateau for that value of $\alpha$.
Finally this value agrees within error bars with the value 2.0, exponent
found experimentally by Beggs and Plenz (\cite{bp03,bp04}).

\begin{figure}
\vskip+1cm
\includegraphics[width=8cm]{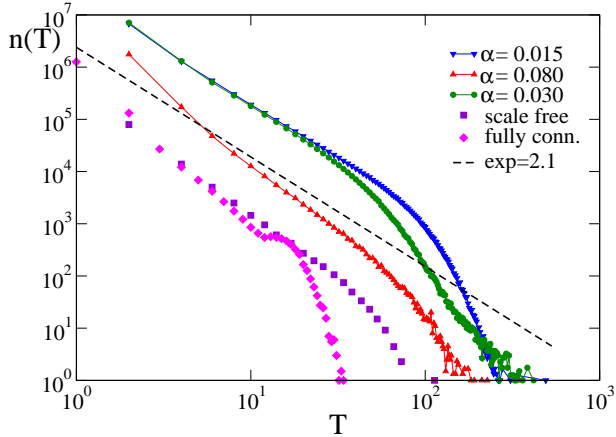}
\caption{Avalanche duration distribution for different
networks with  $p_{in}=0.05$: the scale free network($N=4000$); the fully connected network with
$N_p=50000$; the Apollonian networks for different
values of $\alpha$ (9th generation, $N_p =500$). Data are logarithmically binned.
The dotted line has slope 2.1.}
\label{Fig.7}
\end{figure}

\section{Power spectrum}

The power spectrum of the time signal for the overall electrical activity
can be calculated.
The aim is to compare the scaling behaviour of the numerical spectrum
with the power law observed usually in medical data (\cite{fre,nov}).
For this purpose, the number of active neurons is monitored as function of
time, which recalls the experimental
condition in which electrodes are placed on the scalp in order to
study the patient's spontaneous electrical activity.
In neuronal networks neuronal activity consists in avalanches of
all sizes generated in response to the external stimulus.
Here the unit time is the time for the avalanche to propagate from one
neuron to the next one. The power
spectrum is calculated as the squared amplitude of
the time Fourier transform as
function of frequency, averaged over many initial configurations.
Because of the definition of the numerical time unit, the frequency unit
does not correspond to the experimental one in Hertz.

Fig.8 shows the spectrum for different networks and different
values of $N_p$.
We also show 
the magnetoelectroencephalography (similar to EEG \index{EEG})
obtained from
channel 17 in the left hemisphere of a male
subject resting with his eyes closed, as measured in ref. \cite{nov},
having the exponent 0.795.
For $N_p=0$, i.e. without plasticity adaptation, the spectrum has a
$1/f$ behaviour, characteristic of SOC. For values of $N_p$ different from
zero, but before the $N_{pb}$ plateau, one can distinguish two
different regimes: a power law behaviour with exponent $\beta=0.8\pm 0.1$ at
high frequency, followed by a crossover toward white noise at low frequency.
The difference between $\beta=1$ for $N_p=0$ and $\beta\simeq 0.8$
for higher $N_p$, suggests that plasticity reduces the relevance of small
frequencies in the
power spectrum, in better agreement with experimental EEG
spectra (\cite{fre,nov}).
The stability of the exponent with respect to $\alpha$ has also been verified,
finding  consistently $\beta=0.8\pm 0.1$ at high frequency.
The stability of the spectrum exponent suggests that an universal
scaling characterizes a large class of brain models and physiological signal
spectra for brain controlled activities.
Medical studies of EEG \index{EEG} focus on subtle details of a power spectrum
(e.g. shift
in peaks) to discern between various pathologies. These detailed structures
however live on a background power law spectrum that shows universally an
exponent of about 0.8, as measured for instance in refs.
\cite{fre} and \cite{nov}. A similar
exponent was also detected in the spectral analysis of the stride-to-stride
fluctuations in the normal human gait which can directly be related to
neurological activity (\cite{hau}).
The measured value for
the power spectra exponent is in agreement with the expected relation
$\beta =3-\tau$, being the scaling exponent of the avalanche duration
distribution $\tau >1$ (\cite{jen}).

\begin{figure}
\vskip+1cm
\includegraphics[width=8cm]{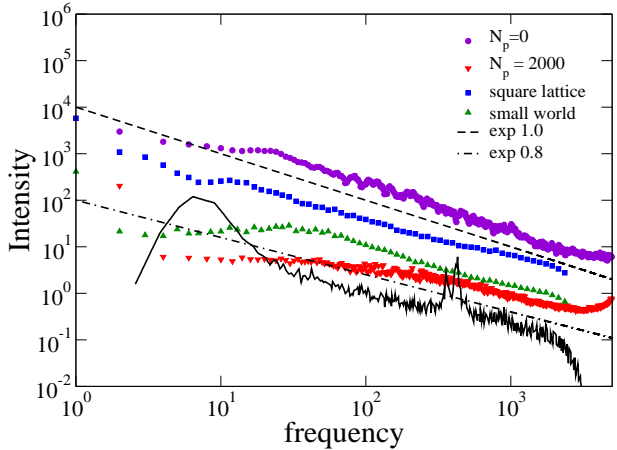}
\caption{Power spectra obtained for different networks:
square lattice ($N = 10^6$, $\alpha=0.03$, $N_p = 10$); 
small world networks ($N = 10^6$, $\alpha= 0.05$, 
$N_p = 1000$, $1 \%$ rewired bonds); Apollonian networks for
different $N_p$ (9th generation, $\alpha=0.020$).
The experimental data
(black line) are from ref. \cite{nov} with frequency in $Hz$. Experimental data
are shifted in order to be in the same frequency range of numerical data.
}
\label{Fig.8}
\end{figure}

The scaling behaviour of the power spectrum can be interpreted in terms
of a stochastic process resulting from the superposition
of multiple inputs taking Gaussian 
distributed random values (\cite{haus}).
The output signal sum of different and uncorrelated
superimposed processes is characterized by a power spectrum with power law
regime, crossing over to white noise at low frequencies and to brown noise to
high frequencies. The low crossover
frequency is related to the inverse of the longest characteristic time
among the superimposed processes. 
$1/f$ noise characterizes a superposition of processes of different
frequencies with similar amplitudes.
In our case the scaling exponent is smaller than
unity, suggesting that processes with high
characteristic frequency are more relevant than processes with low
frequency in the superposition (\cite{haus}).

\section{Discussion}

 Several experimental evidences suggest that the brain behaves as a system
acting at a critical point. This statement implies that the collective
behaviour of the network is more complex than the functioning of the
single components. Moreover, the emergence of self-organized  neuronal
activity, with the absence of a characteristic scale in the response, unveils
similarities with other natural phenomena exhibiting scale-free
behaviour, as earthquakes or solar flares (\cite{easol}). 
For a wide class of these phenomena, self-organized criticality  has
indeed become a successful interpretive scheme.
As in self-organized criticality, the threshold dynamics ensures
time scale separation (slow external drive and fast internal relaxation).
This dynamics leads to criticality and therefore power law behaviour
(\cite{jen}). The model belongs to the class of non-conservative models,
since output neurons can drive charge outside the system. 
However the model presents a number of different features:
The propagation of charge from one neuron to the connected one is non-uniform
 and non-isotropic. Moreover the connectivity network is not static but
dynamically evolves following activity. In this scenario the plastic rules introduce a
homeostatic regulatory mechanism between excitation and inhibition leading to
critical behaviour. The ensemble of these new ingredients is at the
origin of the measured exponents, different from the typical exponents found in
SOC models. Ii is interesting to notice that in fully connected networks the 
excess of very
large avalanches hampers the synaptic depression mechanisms and therefore
alters the self-organized regulation between excitation and inhibition. As a
consequence, supercritical behaviour is observed.

Extensive simulations of this activity dependent 
model are derived for
regular, small world, scale free and fully connected  lattices. The results are
compared with experimental data. The first result is that an optimal value 
of the plasticity  strength $\alpha$ exists with respect to
the pruning  process, optimizing information transmission.
This remark could be interpreted as the evidence of a homeostatic
mechanism between strengthening and weakening processes in the
adaptation of real synapses.
Moreover the avalanche size  and duration distributions exhibit a power law behaviour
with stable exponents 
 compatible with the values experimentally found for neuronal
avalanches. These values appear to be independent of the model parameters
and the specific connectivity network. This universal behaviour is also in
agreement with experimental results, which provide the same exponents for very
different systems (dissociated neurons, cortex slices and networks {\it in vivo}),
evidently characterized by connectivity networks with different complexity.
Solely the fully connected networks consistently exhibit supercritical
behaviour due to the high connectivity level which sustains large avalanche activity.
The stability of the spectrum exponent suggests that a universal
scaling characterizes a large class of brain models and physiological signal
spectra for brain controlled activities.
This work may open new perspectives to study pathological features
of EEG  spectra by including further realistic details into the
neuron and synapse behaviour.

\end{document}